\documentclass[aps,prb,twocolumn,showpacs]{revtex4}
\usepackage{graphicx}

\begin{document}

\title{Self-assembly of quantum dots: effect of neighbor islands on the 
wetting in coherent Stranski-Krastanov growth}

\author{Jos\'e Emilio Prieto}
\email{jeprieto@physik.fu-berlin.de}
\affiliation{Institut f\"ur Experimentalphysik, Freie
Universit\"at Berlin, Arnimallee 14, 14195 Berlin, Germany}

\author{Ivan Markov}
\email{imarkov@ipchp.ipc.bas.bg}
\affiliation{Institute of Physical Chemistry, Bulgarian Academy of Sciences,
1113 Sofia, Bulgaria}
\date{\today}

\begin{abstract}
The wetting of the homogeneously strained wetting layer by dislocation-free 
three-dimensional islands belonging to an array has been studied. 
The array has been simulated as a chain of islands in 1+1 dimensions. 
It is found that the wetting depends on the density of the array, the size 
distribution and the shape of the neighbor islands. Implications for the
self-assembly of quantum dots grown in the coherent Stranski-Krastanov 
mode are discussed.
\end{abstract}


\maketitle

The instability of planar films against the development of coherently 
strain\-ed three-dimensional (3D) islands in highly mis\-matched epitaxy is 
a subject of intense research in recent times due to the optoelectronic
applications of the islands as quan\-tum dots.\cite{Politi} The term 
``coherent Stran\-ski-Krastanov (SK) growth" has been coined for this 
formation of 3D islands that are strained to fit the underlying wetting 
layer but are nearly strainfree near their top and side 
walls,\cite{Eag,Vit} in contrast to the ``classical" SK growth in which 
the lattice mismatch is accommodated by misfit dislocations at the 
interface with the wetting layer.\cite{Matt}

Experimental studies of arrays of coherent 3D islands of 
se\-mi\-con\-duc\-tor ma\-te\-ri\-als have shown surprisingly 
narrow size distributions.\cite{Leo,Moi,Gru,Jia}
This phenomenon, known as self-assembly,\cite{Chris}
is highly desirable as it guarantees a spe\-ci\-fic optical wavelength of the
ar\-ray of quantum dots. The physics of the self-as\-semb\-ly is still not
well understood in spite of several studies.\cite{ter,sch,dobbs,zang}
Priester and Lannoo found that two-dimensional (2D) monolayer-high islands 
have a minimal energy per atom for a certain size and
act as pre\-cur\-sors of the 3D py\-ra\-mi\-dal islands, which
become energetically favored at a smaller volume.\cite{Pri} 
Thus at some critical surface coverage the 2D islands spontaneously transform
into 3D ones preserving a nearly constant volume. The resulting size 
distribution reflects that of the 2D islands which is very narrow. 
This picture has been recently cor\-ro\-bo\-rated
by Ebiko {\it et al.}\cite{Ebi} who found that the volume distribution of
InAs/GaAs self-assembled quantum dots agrees well with the scaling function
characteristic of submonolayer homoepitaxy.
\cite{Amar} Korutcheva {\it et al.}\cite{Kor} and Markov and
Prieto\cite{Marjosem} reached the same conclusion with the exception that the
2D-3D transformation was found to take place through a series of
in\-ter\-mediate states with discretely increasing thickness, in monolayer (ML)
steps, that are stable in separate consecutive
intervals of volume. Khor and Das Sarma arrived to the same conclusion
using Montecarlo simulations.\cite{Khor}

The formation of coherent 3D islands has been discussed within the
framework of the traditional concept of wetting.\cite{Prieto} The wetting
parameter, which accounts for the energetic influence of a crystal B in the
heteroepitaxy of a crystal A on top of it, is defined as\cite{Mar2}
$\Phi  = 1 - E_{\rm AB}/E_{\rm AA}$, 
where $E_{\rm AA}$ and $E_{\rm AB}$ are the energies per atom required to
disjoin a half-crystal A from a like half-crystal A and from an unlike
half-crystal B, respectively.
The mode of growth of a thin film is determined by the difference $\Delta \mu
= \mu (n) - \mu _{\rm 3D}^0$, where $\mu (n)$ and $\mu _{\rm 3D}^0$ are the
chemical potentials of the film (as a function of its thickness $n$) and of the
bulk material A, respectively.\cite{Mar2} 
The chemical potential of the bulk crystal A is given at zero temperature
by $- \phi _{\rm AA}$, the negative of the binding energy of an
atom to the well known kink or half-crystal position. At this position the
atom is bound to a half-atomic row, a half-crystal plane and a half-crystal
block.\cite{Stran}
In the case of a monolayer-thick film of A on the surface of B the chemical
potential of A is given by the analogous energy $- \phi _{\rm AB}$ when the
the underlying half-crystal block of A is replaced by a
half-crystal block of B. Thus $\Delta \mu  = \phi _{\rm AA}
- \phi _{\rm AB}$. In the simplest case of additivity of bond energies this
difference reduces to $E_{\rm AA} - E_{\rm AB}$. Then $\Delta \mu $ is
proportional to $\Phi $, i.e. $\Delta \mu  = E_{\rm AA}\Phi $.\cite{Mar2} 
It follows that the wetting parameter $\Phi $ determines in fact
the mechanism of growth of A on B.\cite{Rudy}

\begin{figure}[h]
\includegraphics*[]{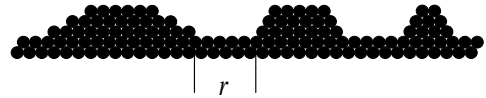}
\caption{\label{model}
Schematic view of an array of islands on a wetting layer. The central island
is surrounded by two islands with different shapes and sizes. The spacing 
between neighbor islands is denoted by $r$, which is a measure of the
density of the array.}
\end{figure}

In the present work we study the behavior of $\Phi $ for islands that
belong to an array. For the coherent SK growth, which
is in fact the formation of dislocation-free 3D islands of A on the same
(strained) material A, it is convenient to define the wetting parameter 
$\Phi $ as the difference of the interaction energies of misfitting and 
non-misfitting 3D islands with the wetting layer.\cite{Prieto} 
We study the effect of the array density and of the size and shape 
distributions of neighbor islands on the wetting parameter $\Phi $ of 
a given island.
We consider an atomistic model in $1+1$ dimensions which can be regarded 
as a cross-section of the real $2+1$ case. An implicit assumption is
that the islands have a compact rather than
a fractal shape and that the lattice misfit is the same in both orthogonal
directions. The 3D islands are represented by linear chains of atoms stacked
one upon the other.\cite{Ratsch} The shape of the islands in our model is
given by the slope of the side walls. The array in the 1+1 dimensional 
space is represented by a row
of 3 or 5 islands on a wetting layer consisting of several monolayers 
(Fig.~\ref{model}). The distance between two neighbor islands is given 
by the number $r$ of vacant atomic positions between the ends of their 
base chains.

In order to simplify the computational procedure, the ``wetting layer" in our
model is in fact composed of several monolayers of the true wetting layer
of the overlayer material A plus several monolayers of the unlike substrate 
material B. This composite wet\-ting layer has the atom
spacing of the substrate material B as is the real case, but for the sake of
simplicity, it has the atom bonding of the overlayer material A. 
This underestimates somewhat the value of $\Phi $ because the A-A bonding 
is weaker than the B-B bonding, but it does not introduce a significant 
error as the energetic influence of the substrate B is screened by 
the true wetting layer A.

To find the equilibrium atomic positions, we make use of 
a simple minimization procedure.~\cite{Marjosem}
The atoms interact through a Morse potential
$V(x)~=~V_0~[e^{-12(x-a)}-2e^{-6(x-a)}]$. 
We calculate the interaction energy of all the atoms as well as its gradient 
with respect to the atomic coordinates, i.e. the forces. Relaxation of the 
system is performed by allowing the atoms to displace in the direction of 
the gradient in an iterative procedure until the forces fall below some
negligible cutoff value. Periodic boundary conditions are applied in the
lateral direction.  We consider only interactions with first neighbors. 

\begin{figure}
\includegraphics*[width=7.0cm]{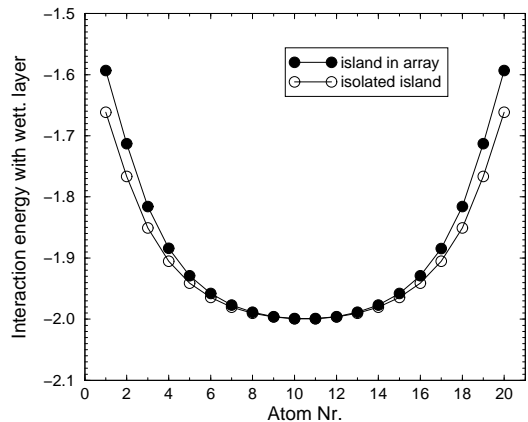}
\caption{\label{evsi}
Distribution of the energy (in units of $V_0$) between the atoms of
the base chain of a 3~ML-high, coherent island of 20 atoms in the base chain, 
and the underlying wetting layer, for a positive misfit of 8\%. Full circles
correspond to an island separated by a distance $r = 5$ from two identical
neighbors, while the empty ones correspond to a reference isolated island.}
\end{figure}

As expected, the edge atoms are found to be more weakly bound to the
underlying wetting layer than the center atoms (Fig.~\ref{evsi}). 
Compared to an isolated island, the edge atoms of an island 
in an array adhere weaker to the substrate. 
This is in fact the essential physical effect exerted by the neighbors  
on a given island: it loses contact with the substrate and
the wetting parameter is increased. We can regard this as the substrate 
becoming stiffer under the influence of the neighbor islands.

\begin{figure}
\includegraphics*[width=7.0cm]{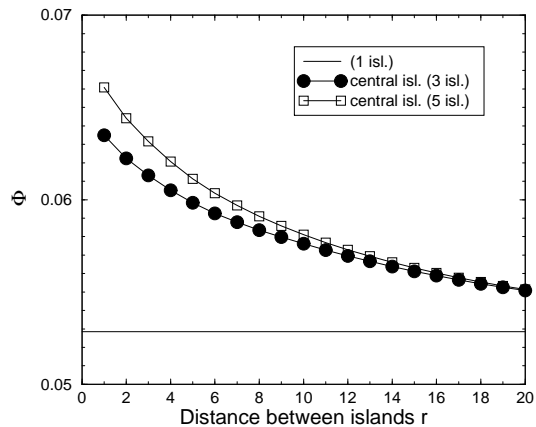}
\caption{\label{phivsr}
Dependence of the wetting parameter of the central island on the distance $r$
between the islands. All islands are 3~ML high and have 20 atoms in the
base chain. The lattice misfit amounts to 7\%. Results for arrays of 3 and 5 
islands are given, as well as for a reference isolated island.}
\end{figure}

The influence of the density of the array is de\-mon\-stra\-ted in 
Fig.~\ref{phivsr}. As expected, the wetting parameter increases with 
increasing array density. The figure also allows us to estimate the 
size of the effect.
At a distance $r$=10 (about 30~nm), neighbors make $\Phi$ increase by 10\%; 
this represents an effective decrease in adhesion $\Delta E_{AB}$ 
of 0.10~$\Phi$~$E_{AA}$. For Ge/Si(100) (desorption energy of Ge: 4~eV), 
this gives about 20~meV, a contribution of the same order as the elastic 
energy per atom (40~meV in this system~\cite{Mar2}) that can significantly 
affect the delicate balance of the energies involved in the growth process: 
diffusion barriers and surface/interface energies.

\begin{figure}
\includegraphics*[width=7.0cm]{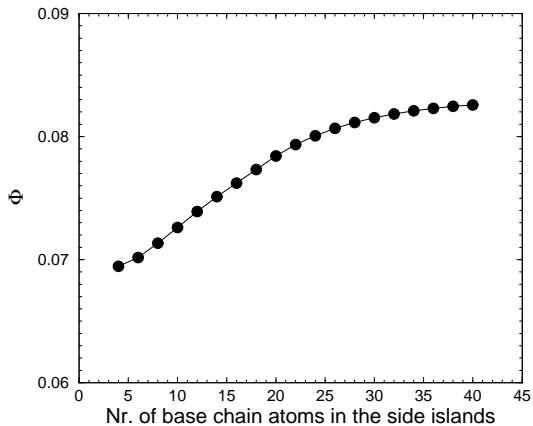}
\caption{\label{phivssize}
Dependence of the wetting parameter of the central island on the size
of the two side islands. These two have equal volumes and are separated 
from the central one by a distance $r$ = 5.
All the islands are 3~ML high, the central one has 20 atoms in the
base chain and the misfit amounts to 7\%.
% and the wetting layer consists of 3~ML, which are allowed to relax.
}
\end{figure}

Figure~\ref{phivssize} shows the wetting parameter of the
central island vs. the size of the side islands. Increasing the volume of the
side islands leads to an increase of the elastic fields around them and to a
further reduction of the bonding between the edge atoms of the central island
and the wetting layer.

\begin{figure}
\includegraphics*[width=7.0cm]{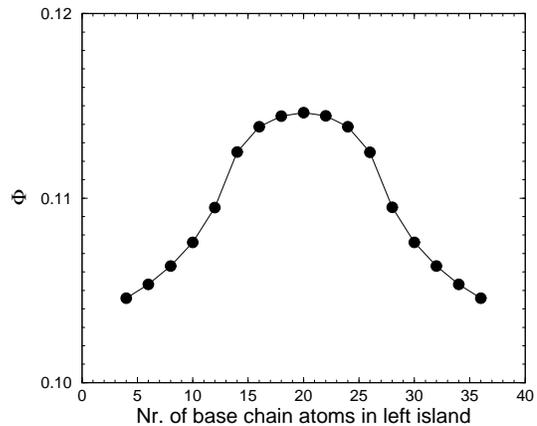}
\caption{\label{phivsnl}
Dependence of the wetting parameter of the central island on the size
distribution of the side islands. The islands are 3~ML thick and are
separated by a distance $r$ = 5; the central one
has 20 atoms in the base chain. The misfit amounts to 8\%.
%The wetting layer consists of 7~ML allowed to relax.
The sum of the volumes of left and right islands is kept constant
and equal to the doubled volume of the central island.
At the center, the three islands have equal volumes.}
\end{figure}

Figure~\ref{phivsnl} demonstrates one further important result, the
effect of the size distribution on the wetting of the islands. It shows the
behavior of the wetting parameter $\Phi $ of the central island as a function
of the number of atoms in the base chain of the left island. 
The sum of the total number of atoms of left and right 
islands is kept constant and precisely equal to the doubled number of the 
central island. 
The facet angle of all three islands is 60$^\circ$. Thus the first (and
the last)
points give the maximum asymmetry in the size distribution of the array, the
left (right) island containing 9 atoms and the right (left) island 
105. All three islands are 3~ML thick.
The point at the maximal wetting corresponds to the monodisperse distribution:
the three islands having one and the same volume of 57 atoms. 
This means that in the case of perfect self-assembly 
of the array, the wetting parameter and therefore the tendency to clustering
display a maximum value.

\begin{figure}
\includegraphics*[width=7.0cm]{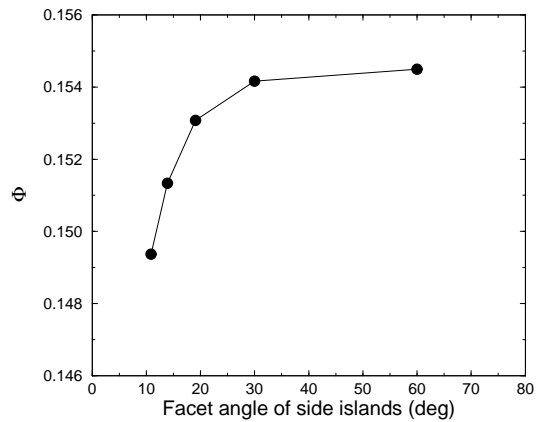}
\caption{\label{phivsang}
Dependence of the wetting parameter of the central island on the shape of the
neighbor islands, as given by the facet angle of their side walls.
The central island has side walls of 60$^\circ$. All islands are 3~ML high,
have 20 atoms in their base chains and are separated by a distance $r$~=~5.
%The wetting layer consists of 3~ML allowed to relax.
The misfit amounts to 8\%.}
\end{figure}

The effect of the shape of the side islands, i.e. their facet angles, on the
wetting parameter of the central island is demonstrated in Fig.~\ref{phivsang}.
The facet angle of the central island is 60$^\circ$. The effect is
greatest when the side islands have the steepest walls. The same result (not
shown) is obtained when the central island has a different shape.
The explanation follows the same line as the one given above: 
islands with larger-angle side walls exert a greater elastic 
effect on the substrate and in turn on
the displacements and the bonding of the edge atoms of the central island. 

When discussing the above results we have to bear in mind that a positive 
value of the wetting parameter implies in fact a tendency of the deposit 
to form 3D clusters instead of a planar film. 
In the case of coherent SK growth the
non-zero wetting parameter is due to the weaker adhesion of the atoms close 
to the island's edges. The presence of other islands, particularly with 
large angle facets, in the near vicinity enhances the effect.
The transformation of 2D, monolayer-high islands into
bilayer islands takes place by detachment of atoms from the edges and their
subsequent jumping and nucleating on the top island's surface.\cite{Stmar} 
This edge effect hints at the influence of the lattice misfit on the
rate of second layer nucleation and in turn on the kinetics of the 2D-3D
transformation.\cite{Fil,Lin} The presence of neighbor islands favors
the formation of 3D clusters and their further growth. For the self-assembled
monodisperse population, the highest tendency to clustering is found.

We can regard the flatter islands in our model (11$^\circ$ facet angle) as
the ``hut" clusters discovered by Mo {\it et al.},\cite{Mo} and the
clusters with 60$^\circ$ facet angle as the ``dome" clusters. It is well
known that clusters with steeper side walls relieve the strain more
efficiently than flatter clusters~\cite{Mich} (a planar film, the 
limiting case of the flat islands, does not relieve strain at all). 
We see that large-angle
facet islands affect more strongly the growth of the neighbor 
islands, leading to a more narrow size distribution.

From our results, a self-assembled population of quantum dots with 
highest density is expected at comparatively low tem\-pe\-ra\-tures such 
that the critical wetting layer thickness for 3D islanding approaches an 
integer number of monolayers. In InAs/GaAs quantum dots the reported values 
of the critical thickness are found to vary from 1.2 to 2~ML.\cite{Pol} The
critical wetting layer thickness is given by an integer number $n$ of
monolayers plus the product of the 2D island density and the critical volume
$N_{12}$ in the ($n$+1)-th~ML. 
The 2D island density increases steeply with decreasing 
temperature.\cite{dobbs} 
In such a case a dense population of 2D islands will overcome 
simultaneously the critical size $N_{12}$ to produce bilayer islands. 
The value of $N_{12}$ will be slightly reduced when neighbor
islands are present due to the increase of $\Phi$.\cite{Prieto}
Regions of high adatom concentrations will favor the highest degree of 
self-assembly and, due to the larger elastic forces present, are also 
likely to promote the spatial ordering of the islands, possibly 
extending to less dense regions and leading to self-organized arrays.
Islands will thus interact with each other from the very beginning of the 
2D-3D transformation and will give rise to the maximum possible wetting 
parameter and, in turn, to islands with large-angle facets and a narrow size 
distribution. This is in agreement with the observations of self-assembled
Ge quantum dots on Si(001).\cite{Le} At 700$^\circ$C a population of islands
with a concentration ranging from $10^7$ to $10^8$ cm$^{-2}$
is obtained. The islands have the shape of truncated square pyramids with 
their side wall facets formed by (105) planes (inclination angle of
about 11$^\circ$). The size distribution of the islands is quite broad. At
550$^\circ$C a population of islands with an areal density of the order of
$10^9$ to $10^{10}$ cm$^{-2}$ is observed. The islands have larger-angle 
(113) facets and their size distribution is much more narrow. 

In summary, the presence of neigh\-bor is\-lands de\-creases the 
wetting of the substrate (in this case the wetting layer) by the 3D islands. 
The wetting decreases with increasing array density and facet angle 
of the neighbor islands.
The wetting parameter displays a maximum (implying 
a minimal wetting) when the array shows a monodisperse size distribution. 
We expect an optimum self-assembled islanding
at low temperatures such that the 2D-3D transformation takes place at the
highest possible island density.

J. E. P. gratefully acknowledges financial support from the 
Alexander-von-Humboldt Stiftung and the Spanish MEC 
(grant No. EX2001 11808094).

\end{document}